\documentclass[conference]{IEEEtran}
\IEEEoverridecommandlockouts

\usepackage{cite}
\usepackage{amsmath,amssymb,amsfonts}
\usepackage{graphicx}
\usepackage{textcomp}
\usepackage{xcolor}
\usepackage{booktabs}
\usepackage{comment}
\usepackage{multirow}
\usepackage{pifont}
\newcommand{\xmark}{\ding{55}}%
\def\BibTeX{{\rm B\kern-.05em{\sc i\kern-.025em b}\kern-.08em
    T\kern-.1667em\lower.7ex\hbox{E}\kern-.125emX}}
\usepackage[caption=false,font=footnotesize]{subfig}
\usepackage{enumitem}
\usepackage{xspace}
\newcommand{\eg}{\textit{e.g.},\xspace}
\newcommand{\ie}{\textit{i.e.},\xspace}
\usepackage{algorithm}
\usepackage[noend]{algpseudocode}
\usepackage[hidelinks]{hyperref}
\usepackage{balance}

\usepackage{tikz}
\newcommand\publishedtext{%

\footnotesize This is the author's accepted version of the article. The final version published by IEEE is J. Suárez-Varela, and A. Lutu, ``Uncovering Issues in the Radio Access Network by Looking at the Neighbors,`` IEEE International Conference on Machine Learning for Communication and Networking, 2025, pp. TBP, doi: TBP.}  
\newcommand\copyrighttext{%
\footnotesize \textcopyright 2025 IEEE. Personal use of this material is permitted. Permission from IEEE must be obtained for all other uses, in any current or future media, including reprinting/republishing this material for advertising or promotional purposes, creating new collective works, for resale or redistribution to servers or lists, or reuse of any copyrighted component of this work in other works.}
\newcommand\copyrightnotice{%
\begin{tikzpicture}[remember picture,overlay]
\node[anchor=north,yshift=0pt] at (current page.north) {\fbox{\parbox{\dimexpr\textwidth-\fboxsep-\fboxrule\relax}{\publishedtext}}};
\node[anchor=south,yshift=10pt] at (current page.south) {\fbox{\parbox{\dimexpr\textwidth-\fboxsep-\fboxrule\relax}{\copyrighttext}}};
\end{tikzpicture}%
}

\usepackage{xspace}
\newcommand{\anemon}{\mbox{c-ANEMON}\xspace}

\makeatletter
\renewcommand\subsubsection{\@startsection{subsubsection}{3}{\z@}%
  {1ex plus 1ex minus .2ex}
  {1ex}
  {\normalfont\normalsize\itshape}} 
\makeatother

\begin{document}

\title{Uncovering Issues in the Radio Access Network \\ by Looking at the Neighbors
}

\author{
\IEEEauthorblockN{Jos\'e Su\'arez-Varela}
\IEEEauthorblockA{\textit{Telef\'onica Research} \\
jose.suarez-varela@telefonica.com}
\and
\IEEEauthorblockN{Andra Lutu}
\IEEEauthorblockA{\textit{Telef\'onica Research} \\
andra.lutu@telefonica.com}
}

\maketitle

\copyrightnotice\vspace*{-9pt}

\begin{abstract}
Mobile network operators (MNOs) manage Radio Access Networks (RANs) with massive amounts of cells over multiple radio generations (2G-5G). To handle such complexity, operations teams rely on monitoring systems, including anomaly detection tools that identify unexpected behaviors. In this paper, we present \anemon, a \textit{Contextual ANomaly dEtection MONitor for the RAN} based on Graph Neural Networks (GNNs). 
Our solution captures spatio-temporal variations by analyzing the behavior of individual cells in relation to their local neighborhoods, enabling the detection of anomalies  that are independent of external mobility factors. This, in turn, allows focusing on anomalies associated with network issues (\eg misconfigurations, equipment failures). We evaluate \anemon using real-world data from a large European metropolitan area (7,890 cells; 3~months). 
First, we show that the GNN model within our solution generalizes effectively to cells from previously unseen areas, suggesting the possibility of using a single model across extensive deployment regions. Then, we analyze the anomalies detected by \anemon through manual inspection and define several categories of long-lasting anomalies (6+ hours). Notably, 45.95\% of these anomalies fall into a category that is more likely to require intervention by operations teams.
\end{abstract}

\section{Introduction}

Large Mobile Network Operators (MNOs) provide countrywide connectivity to millions of subscribers, relying on hundreds of thousands of cells distributed across the Radio Access Network (RAN) and supporting the coexistence of multiple radio access generations (\eg 2G-5G). To deal with such complexity, operations teams rely on commercial monitoring solutions that generate alarms with varying levels of severity~\cite{NokiaNetAct,netscout5g}, and a ticketing system is in place for the management of critical incidents (either raised by these monitoring solutions or other sources). However, some issues still remain undetected, leading to temporary service degradation and---in the worst case---incidents reported by customers.

Anomaly detection tools facilitate the early recognition of irregular events, which may stem from different root causes, such as equipment malfunctions, misconfigurations, cyberattacks, or abrupt variations in the traffic demand~\cite{fernandes2019comprehensive, wang2021machine}. As a result, these tools allow for the prompt identification and troubleshooting of network issues that otherwise might severely affect the end-user experience.

With the advent of BigData and Machine Learning (ML), the mobile network community has put a significant effort in developing automatic data-driven methods that can efficiently detect anomalies in the RAN~\cite{iyer2017automating, liu2009aldo, sun2024spotlight}. Also, traffic anomaly detection has been widely studied in the past, with some popular works that apply Principal Component Analysis (PCA) to detect anomalies on network-wide traffic~\cite{lakhina2004diagnosing}. 
All these methods focus on learning the normal distributions of Key Performance Indicators (KPIs) on individual elements (\eg radio cells), and then using the learned distributions to detect anomalous deviations during inference.  This typically requires training and maintaining different models for each cell, or groups of cells (\eg at the site level). Moreover, cellular networks are known to exhibit marked spatio-temporal patterns as a result of the mobility of users connected to the network~\cite{wang2018spatio,zhang2018long}, which leads to noisy and highly dynamic KPI time series that are challenging to model. 

In this paper, we put forward the idea that anomalies can be detected by monitoring the state of cells, finding other cells with similar past activity (\ie context cells), and detecting when a cell deviates from its context. Consequently, we focus on contextual anomaly detection, which can complement existing RAN anomaly detection solutions by finding other types of anomalies undetected by these methods~\cite{dimopoulos2017detecting}. 

Expanding on this concept, we present \anemon, a \textit{\mbox{Contextual} ANomaly MONitor} for the RAN based on graph representations and Deep Learning. Our solution integrates a custom model based on Graph Neural Networks (GNN)~\cite{battaglia2018relational} that efficiently processes the information in the local neighborhood of cells. A key feature of our solution is that the model learns from the collective experience of all cells in the training dataset and it automatically adapts its predictions based on the local context (\eg local mobility patterns).
Then, we use this model to detect anomalous deviations in production.
As a result, \anemon focuses on detecting anomalies that are independent of external mobility factors (\eg massive events, weather conditions). Additionally, the GNN-based model within our solution implements a novel architecture that enhances the interpretability of the context, which we then leverage to improve the robustness of our solution and minimize false alarms. This is crucial to facilitate that operations teams can focus on relevant issues, which ultimately contributes to OPEX savings.

In our evaluation, we use real-world data from a top-tier MNO in a densely populated metropolitan area in Europe (7,890 cells; 3 months). 
First, we test the generalization capabilities of the proposed GNN-based model (Sec.~\ref{subsec:training-predictor}), and show that it achieves accurate predictions even when we apply it to cells of another area not included in the training dataset. 
This suggests the possibility to train the solution in a reduced trial area, and then apply it for anomaly detection in larger deployment regions, where cells have similar features (\eg same frequency bands, antenna kits, deployment density). 
Second, we delve into the anomalies detected by \anemon through manual inspection (Sec.~\ref{subsec:evaluation-anomalies}). 
Particularly, we focus on a subset of anomalies extending over longer time periods (6+ hours) and define several classes. As a result, we identify a class that accounts for 45.95\% of the anomalies. This class is especially relevant from an operational perspective, as it includes cases that are more likely to require manual intervention. Lastly, we test our solution on a massive event to evaluate its robustness against extreme mobility patterns (Sec.~\ref{subsec:robustness-evaluation}). The results show that \anemon effectively adapts to the dynamic conditions of the scenario and remains unaffected by the external mobility factor.

\section{Related Work}

Network anomaly detection is a well-studied research area, with a wide range of solutions applied over multiple use cases and network scenarios~\cite{fernandes2019comprehensive, wang2021machine}. We summarize below the main classes of anomaly detection methods relevant to this work.

\subsection{Time series anomaly detection}
Some anomaly detection methods rely on time-series representations (\eg for traffic analysis) using data-driven methods ranging from ARIMA-based models~\cite{yaacob2010arima} to more recent Deep Learning models based on Recurrent Neural Networks (\eg LSTMs~\cite{kilinc2022jade, trinh2019detecting}). These types of solutions train ML models to predict historical KPI data over individual cells in the RAN, which are then applied to detect deviations with respect to the expected behavior~\cite{kilinc2022jade, trinh2019detecting,sun2024spotlight}. Additionally, other methods are based on the reconstruction error, for which they use models based on variational autoencoders, including architectures with Convolutional Neural Networks~\cite{gonzalez2022dc}, Transformers~\cite{tuli2022tranad}, or generative models~\cite{sun2024spotlight}. While these state-of-the-art time series anomaly detection methods have shown strong performance for some applications in the RAN~\cite{trinh2019detecting,gonzalez2022dc,sun2024spotlight}, they present some shortcomings from the network operations perspective: these methods are not designed to differentiate between anomalies caused by external factors (\eg user mobility between cells) and those arising from network-related issues (\eg misconfigurations, equipment failures).

\subsection{Contextual anomaly detection}
These anomaly detection methods strive to capture anomalies between elements with some type of relationship, which is often unknown beforehand~\cite{bhuyan2013network}. A typical case is fraud detection in banking operations~\cite{hilal2022financial}, where it is essential not only to analyze transactions at the customer level but also to contextualize them in relation to other customers with similar profiles. In the literature, we can find some methods proposing contextual anomaly detection in time series data~\cite{fernandes2019comprehensive}. These methods first look for sets of elements with similar past activity (\ie the contexts) and then detect deviations of specific elements with respect to their context (\ie the anomalies). For example,~\cite{dimopoulos2017detecting} performs contextual anomaly detection to find issues on end-users in broadband networks, or~\cite{hayes2014contextual} seeks to detect anomalies in sensor monitoring data.

Some recent works propose using GNNs~\cite{battaglia2018relational} for contextual anomaly detection. For example, in~\cite{deng2021graph} the authors perform anomaly detection in multivariate time series from a distributed sensor network in a water treatment plant. Likewise,~\cite{latif2023detecting} proposes a GNN-based method for anomaly detection on network-wide traffic in ISP networks.

In the context of RAN anomaly detection, mobility trends and other external factors cannot be captured by looking individually at the cell-level behavior, but we typically need to consider the evolution of KPIs in nearby cells~\cite{zhang2018long}. This motivates the use of contextual anomaly detection methods that can distill anomalies independent of external mobility factors and, consequently, focus on issues related to the network (\eg misconfigurations). To the best of our knowledge, \anemon is the first solution that leverages the benefits of GNNs to perform contextual anomaly detection in the RAN. Specifically, our solution was carefully designed to account for key aspects of RAN anomaly detection, which we summarize in its main design principles---described in Sec.~\ref {subsec:design-principles}.

\section{Proposed Solution}
In this section, we introduce \anemon, our unsupervised anomaly detection monitor that leverages the use of graphs to represent cells in the RAN deployment. 
Our solution incorporates a custom GNN-based model (referred to as \textit{Contextual Predictor}) that processes information from the neighborhood of cells in an effective manner. Lastly, we use this model to detect anomalies in KPIs at the cell level.

\subsection{Design principles}\label{subsec:design-principles}
We describe below the main drivers behind the design of \anemon:\\
\noindent$\bullet$ \textbf{Context-based anomalies:} we aim to focus on anomalies related to network issues, such as misconfigurations or failures on network equipment. To this end, our contextual solution considers the evolution of KPIs in neighboring cells. This enables to capture mobility patterns in the local area of cells and detect only anomalies that are independent of them. With this, we aim to complement existing anomaly detection methods that are primarily based on analysis at the individual cell level~\cite{iyer2017automating, sun2024spotlight}.

\noindent$\bullet$ \textbf{Early detection:} \anemon is unsupervised and can operate on time series at any time granularity. This allows for early detection of new issues appearing in the network, before they impact the customers’ Quality of Experience. 

\noindent$\bullet$ \textbf{Robust and easy to maintain}: Thanks to the use of a GNN-based model, our solution generalizes robustly to cells from different deployment areas, as shown in Sec.~\ref{subsec:training-predictor}. This allows for training a single GNN model with data from a reduced area, and then deploy it over larger regions. In contrast, state-of-the-art solutions are mostly designed to train independent models for each element (\ie one per cell)~\cite{tuli2022tranad, trinh2019detecting, gonzalez2022dc}.

\noindent$\bullet$ \textbf{Interpretable by design}: We introduce a novel mechanism  within the architecture of our GNN-based model that permits to directly interpret the context captured for a specific sample. This allows us to identify samples with unreliable predictions, which we pre-filter during the anomaly detection process to minimize the false alarm rate (Sec.~\ref{subsubsec:anomaly-detector}).

\subsection{Architecture} \label{subsec:architecture}
\begin{figure}[t!]
    \centering
    \includegraphics[width=1.0\columnwidth]{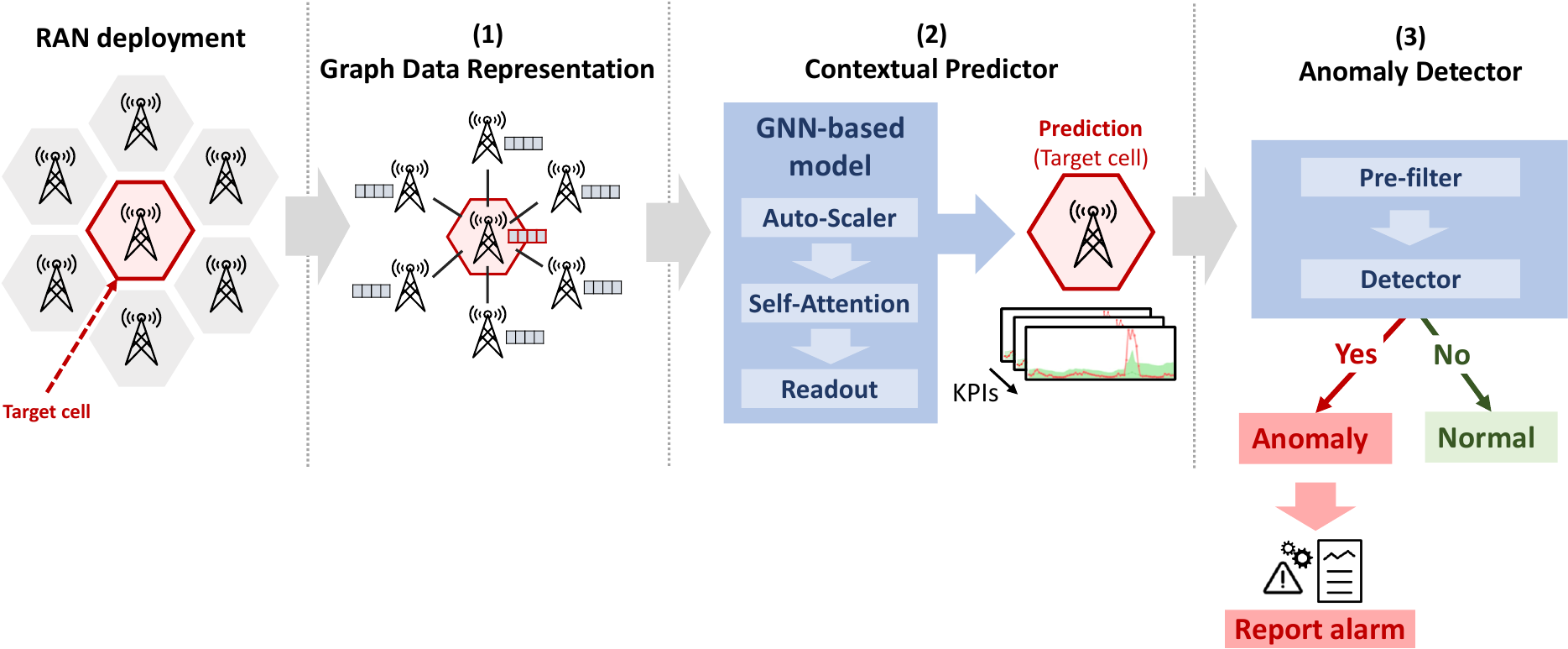}
    \caption{Schematic representation of \anemon.}
    \label{fig:solution-overview}
\end{figure}

Figure~\ref{fig:solution-overview} shows a schematic representation of \anemon, which can be divided into three main components: (1)~the \textit{Graph Data Representation}, (2)~the \textit{\mbox{Contextual} \mbox{Predictor}}, and (3)~the \textit{Anomaly Detector}. The following subsections describe the design of these three components.

\subsubsection{\textbf{Graph Data Representation:} \textnormal{[Fig.~\ref{fig:solution-overview}-(1)]}} \label{subsubsec:data-representation}
\noindent We consider a graph representation where each cell is represented as a node, and edges connect nodes (\ie cells) based on some predefined criteria (\eg distance). Graph edges define the potential relationships, which limit the set of cells that may be considered in the context of a specific cell. Particularly, in each graph sample we define a target cell on which we run the anomaly detector, and we add the $k$ nearest neighbors as connected nodes. In our scenario, each site (or base station) typically hosts multiple cells oriented to different directions and operating in various frequency bands. We represent these cells as independent nodes in the graph, each with its associated KPIs.

Formally, we have a graph $\mathcal{G}=\left(\mathcal{N},\mathcal{E}\right)$, where 
$\mathcal{N}$ represents the set of nodes, corresponding to individual cells, and $\mathcal{E}$ the set of edges, which connect the target cell $(tar)$ with the $k$ neighboring cells, denoted as $\mathcal{N}_{(tar)}$. Each cell $v_i\in\mathcal{N}$ is represented by a vector $\mathbf{c}_i = \big[\mathbf{x}_{t-T:t}^{(i)}, \mathbf{m}_i\big]$, where \mbox{$\mathbf{x}_{t-T:t}^{(i)} = \{x_i(t-T), \dots, x_i(t)\}$} is a sequence of KPI measurements over the \textit{Context Window} $[t-T, \dots, t]$, and $\mathbf{m}_i$ includes some information about the cell (\eg antenna type, frequency band). We denote the information of the target cell as $\mathbf{c}_{t-T:t}^{(tar)}$, and the information of the neighboring cells as $\mathbf{c}_{t-T:t}^{(neigh)} = \{c_{t-T:t}^{(j)} \mid j \in \mathcal{N}_{(tar)}\}$. The latter set contains the context information obtained from the neighbors.

In addition to the Context Window,  we define the \textit{Prediction Window} $[t+1,…,t+L]$, in which the Contextual Predictor (Sec.~\ref{subsubsec:contextual-predictor}) generates its estimates.

\algnewcommand\algorithmicforeach{\textbf{for each}}
\algdef{S}[FOR]{ForEach}[1]{\algorithmicforeach\ #1\ \algorithmicdo}
\renewcommand{\algorithmicrequire}{\textbf{Input:}}
\renewcommand{\algorithmicensure}{\textbf{Output:}}
\begin{algorithm}[!t]
\caption{GNN-based Contextual Predictor}\label{alg:contextual-predictor}
\begin{algorithmic}[1]
\Require $\mathcal{N}, \mathcal{E}, tar$
\Ensure $\mathbf{\hat{x}}_{t+1:t+L}^{(tar)},\: \mathbf{\hat{\sigma}}_{t+1:t+L}^{(tar)}$,\: $(\alpha_j, sc_j, sh_j) \, \forall j \in \mathcal{N}_{(tar)}$
\vskip 0.1cm 
\State \textbf{Auto-Scaler module:}
\ForEach{$c_j \in \mathcal{N}_{(tar)}$}
\State $[\mathbf{x}_{t-T:t}^{(j)}, \mathbf{m}_j] \gets \mathbf{c}_j$ \Comment{Extract time series \& cell info}
\State $sc_j, sh_j \gets \text{ScaleNN}(c_j, c_{tar})$ \Comment{Scale \& shift}
\State $\mathbf{xs}_j \gets \mathbf{x}_{t-T:t}^{(j)} \cdot sc_j + sh_j $ \Comment{Apply transformation}
\State $\mathbf{c}_j \gets [\mathbf{xs}_j, \mathbf{m}_j] $ \Comment{Update cell state}
\EndFor
\vskip 0.1cm 
\State \textbf{Self-Attention module:}
\State $\mathbf{hq}_{tar} \gets \text{QueryEmbed}(c_{tar})$ \Comment{Embedding target cell}
\State $d_k \gets \text{len}(\mathbf{hq}_{tar})$
\ForEach{$c_j \in \mathcal{N}_{(tar)}$}
    \State $\mathbf{hk}_j \gets \text{KeyEmbed}(c_j)$ \Comment{Embedding neighboring cell}
    \State $\alpha_j \gets \frac{\sum (\mathbf{hk}_j \odot \mathbf{hk}_j)}{\sqrt{d_k}}$ \Comment{Attention coefficients}
    \State $\alpha_j \gets \text{Softmax}(\alpha_j)$ \Comment{Normalize}
\EndFor
\vskip 0.1cm 
\State \textbf{Readout module:}
\State // Transformation in Prediction Window $[t+1, ..., t+L]$
\ForEach{$c_j \in \mathcal{N}_{(tar)}$}
\State $\mathbf{xs}_{t+1:t+L}^{(j)} \gets \mathbf{x}_{t+1:t+L}^{(j)} \cdot sc_j + sh_j $
\EndFor
\State // Compute estimate (weighted mean)
\State $\mathbf{\hat{x}}_{t+1:t+L}^{(tar)} \gets \sum_{j \in \mathcal{N}_{(tar)}} \left(\alpha_j \cdot \mathbf{xs}_{t+1:t+L}^{(j)} \right)$
\State // Weighted standard deviation
\State $v_1 \gets \sum_{j \in \mathcal{N}_{(tar)}} \left(\alpha_j\right); \quad v_2 \gets \sum_{j \in \mathcal{N}_{(tar)}} \left(\alpha_j^2\right)$
\vskip 0.1cm
\State $\mathbf{\hat{\sigma}}_{t+1:t+L}^{(tar)} \gets \sqrt{\sum_{j \in \mathcal{N}_{(tar)}} \left(\frac{\alpha_j \cdot \left(\mathbf{xs}_{t+1:t+L}^{(j)} - \mathbf{\hat{x}}_{t+1:t+L}^{(tar)}\right)^2}{v1-(v2/v1)}\right)}$
\end{algorithmic}
\end{algorithm}
\setlength{\textfloatsep}{0.2cm}

\subsubsection{\textbf{Contextual Predictor:} \textnormal{[Fig.~\ref{fig:solution-overview}-(2)]}} \label{subsubsec:contextual-predictor}
\noindent Algorithm~\ref{alg:contextual-predictor} shows a detailed description of the design of our GNN-based Contextual Predictor. It can be divided into three main modules: ($i$)~the \textit{Auto-Scaler}, ($ii$)~the \textit{Self-Attention}, and ($iii$)~the \textit{Readout}.

In the Auto-Scaler module (lines 1-6), the model takes the time series of the neighbors over the Context Window $\mathbf{x}_{t-T:t}^{(j)}$ and applies a linear transformation that involves scaling and shifting (line 5). The scale and shift parameters are computed by a neural network $ScaleNN$ that takes the state of each neighboring cell $c_j$ and combines it with the state of the target cell $c_{tar}$ (line 4). We show in Sec.~\ref{subsec:training-predictor} that the addition of this Auto-Scaler module leads to better performance for the Contextual Predictor.

Then, the model applies the Self-Attention module (lines 7-13), which computes an attention coefficient $\alpha_j$ for each neighboring cell $c_j \in \mathcal{N}_{(tar)}$. For that, we use a self-attention mechanism similar to that of Transformers \cite{vaswani2017attention}. First, the model generates query and key embeddings (lines 8 and 11) and then combines them using a dot product divided by a scaling factor $d_k$ (line 12). The attention coefficients are finally normalized using a Softmax layer (line 13).

Lastly, the Readout module (lines 14-22) generates estimates for the target cell over the Prediction Window $\mathbf{\hat{x}}_{t+1:t+L}^{(tar)}$. Particularly, we build the predictions as a weighted linear combination of the transformed time series of neighboring cells $\mathbf{xs}_{t+1:t+L}^{(j)}$ (lines 15-19). This novel approach differs from traditional Graph Attention Networks (GATs)~\cite{velivckovic2017graph} by enabling the use of attention coefficients ($\alpha_j$) as direct measures of the context captured by the model.\footnote{As shown in Sec.~\ref{subsec:training-predictor}, this modification in the model does not compromise the performance of the Contextual Predictor in our use case.} We then leverage this interpretability feature to avoid potential false alarms in the Anomaly Detector (Sec.~\ref{subsubsec:anomaly-detector}). Additionally, our model estimates the standard deviation $\mathbf{\hat{\sigma}}_{t+1:t+L}^{(tar)}$ (lines 20-22), enabling the use of adaptive confidence intervals in the anomaly detection.

\subsubsection{\textbf{Anomaly Detector:} \textnormal{[Fig.~\ref{fig:solution-overview}-(3)]}} \label{subsubsec:anomaly-detector}
This module takes the estimates of the Contextual Predictor $\mathbf{\hat{x}}_{t+1:t+L}^{(tar)}$ as input and compares them with the actual measurements in the target cell $\mathbf{x}_{t+1:t+L}^{(tar)}$ to determine whether there is an anomaly. It is divided into two main modules: ($i$)~the \textit{Pre-Filter}, and ($ii$)~the \textit{Detector}. 
\vspace{0.2cm}

\noindent \textit{($i$) Pre-Filter}: A key requirement when designing anomaly detection solutions is to reduce the false alarm rate to avoid unneeded manual inspections by the operation teams. Hence, we apply a pre-filter that allows us to discard samples with unreliable predictions by exploiting the interpretability of the Contextual Predictor. We focus on two main aspects based on expert knowledge:

\noindent$\bullet$ \textit{Context size}: The larger the set of cells captured by the model's context, the more robust are potentially the predictions (\ie more information). To measure the size of the effective context, we compute the entropy of the attention coefficients generated by the model $[\alpha_j \in \mathcal{N}_{(tar)}]$ and use it as a proxy for the amount of cells contributing to the context. We then define a configurable parameter $\lambda$ and filter out samples with low entropy values $[entropy(\alpha_j \in \mathcal{N}_{(tar)}) < \lambda]$, which correspond to cases where the attention concentrates on a small subset of cells.

\noindent$\bullet$ \textit{Context reliability}: We check that cells contributing to the context keep a similar historical trend to that of the target cell. For that, we exploit the scale ($sc_j$), shift ($sh_j$) and attention ($\alpha_j$) coefficients and apply them backwards, over the Context Window \mbox{$[t-T, \dots, t]$}. Specifically, we define the normalized historical error $\mathbf{h_{err}}$ in Algorithm~\ref{alg:historical-error}. Note that this produces a vector with errors along the Context Window, hence we define a configurable percentile ($P$) and an associated threshold ($\gamma$), and discard those samples where $\text{pct}_P (\mathbf{h_{err}}) > \gamma$.

\begin{algorithm}[!t]
\caption{Normalized Historical Error}\label{alg:historical-error}
\begin{algorithmic}[1]
\Require $\mathcal{N}, \mathcal{E}, tar, (\alpha_j, sc_j, sh_j) \, \forall j \in \mathcal{N}_{(tar)}$
\Ensure $\mathbf{h_{err}}$ 
\State // Apply transformation in Context Window $[t-T, ..., t]$
\ForEach{$c_j \in \mathcal{N}_{(tar)}$}
\State $\mathbf{xs}_{t-T:t}^{(j)} \gets \mathbf{x}_{t-T:t}^{(j)} \cdot sc_j + sh_j $
\EndFor
\State // Compute historical weighted mean
\State $\mathbf{\hat{x}}_{t-T:t}^{(tar)} \gets \sum_{j \in \mathcal{N}_{(tar)}} \left(\alpha_j \cdot \mathbf{xs}_{t-T:t}^{(j)} \right)$
\vskip 0.1cm
\State // Weighted standard deviation
\State $v_1 \gets \sum_{j \in \mathcal{N}_{(tar)}} \left(\alpha_j\right); \quad v_2 \gets \sum_{j \in \mathcal{N}_{(tar)}} \left(\alpha_j^2\right)$
\vskip 0.1cm
\State $\mathbf{\hat{\sigma}}_{t-T:t}^{(tar)} \gets \sqrt{\sum_{j \in \mathcal{N}_{(tar)}} \left(\frac{\alpha_j \cdot \left(\mathbf{xs}_{t-T:t}^{(j)} - \mathbf{\hat{x}}_{t-T:t}^{(tar)}\right)^2}{v1-(v2/v1)}\right)}$
\vskip 0.1cm
\State // Historical error
\vskip 0.1cm
\State $\mathbf{h_{err}} \gets \frac{|\mathbf{\hat{x}}_{t-T:t}^{(tar)} - \mathbf{x}_{t-T:t}^{(tar)}|}{\mathbf{\hat{\sigma}}_{t-T:t}^{(tar)}}$
\end{algorithmic}
\end{algorithm}

\vspace{0.2cm}
\noindent \textit{($ii$) Detector}:
For those samples that pass the criteria of the Pre-Filter, we consider the predictions $\mathbf{\hat{x}}_{t+1:t+L}^{(tar)}$ and standard deviations $\mathbf{\hat{\sigma}}_{t+1:t+L}^{(tar)}$ produced by the Contextual Predictor (see algorithm~\ref{alg:contextual-predictor}), and compute a vector of anomaly scores $\mathbf{S}_{t+1:t+L}^{(tar)}$. This vector uses a formula similar to that of the historical error (see Algorithm~\ref{alg:historical-error}), but applied over the Prediction Window. Specifically, we define it as: 
$$\mathbf{S}_{t+1:t+L}^{(tar)} = \left(|\mathbf{\hat{x}}_{t+1:t+L}^{(tar)} - \mathbf{x}_{t+1:t+L}^{(tar)}|\right) / \mathbf{\hat{\sigma}}_{t+1:t+L}^{(tar)}$$

Lastly, we define a sensitivity threshold ($\delta$), and the Anomaly Detector labels as anomalous those time samples where $\mathbf{S}_{t+1:t+L}^{(tar)} > \delta$.

\section{Evaluation}

In this section, we train and evaluate \anemon using real-world data from a MNO. Next, we characterize different types of anomalies detected by our solution. Lastly, we test \anemon over an extreme mobility event to assess its robustness to external factors. 

\subsection{Dataset}
We use real-world data from a RAN monitoring platform of a major MNO in a European country. Our dataset contains hourly samples of KPIs from all the 4G/5G cells deployed in a metropolitan area, covering a total of 7,890 cells deployed in 1,300+ sites, for a period of three months (Feb.~1-Apr.~30, 2024). For our experiments, we collect time series of Physical Resource Block (PRB) utilization at the cell level, which includes a total of 15+ million data points.

In the dataset, some cells exhibit high rates of missing data (\eg due to issues in data collection and storage); hence we discard cells with $>$50\% of missing data, which accounts for $\approx$17\% of the cells. For the remaining cells, we apply a data imputation mechanism that repeats values from the most recent available data points in the time series. This helps preserve the trend in the time series and avoid the representation of missing data as outlier data points in the Contextual Predictor.

To train and evaluate our solution, we generate graph samples for each target cell over an hourly sliding window (see Sec.~\ref{subsubsec:data-representation}). Each sample contains 1-week context time series of PRB utilization (\ie $T$=167 hours) for the $k$=200 nearest cells by physical distance. Also, we include the time series of the neighboring cells during the Prediction Window, which is set to $L$=24 hours.

Lastly, we divide our dataset into training, validation, and test by selecting cells from three distinct regions in the analyzed metropolitan area. These splits account for 70\%, 10\%, and 20\% of the cells in the metropolitan area, respectively. This approach ensures that the validation and test datasets include cells unseen during the training phase, allowing us to evaluate the generalization capabilities of the proposed GNN-based Contextual Predictor across different deployment areas.

\subsection{Training the Contextual Predictor} \label{subsec:training-predictor}

We first train the Contextual Predictor on the training dataset. Due to the skewed data distribution of PRB utilization, we apply logarithmic normalization to mitigate the influence of higher values. Additionally, we use a Negative Log Likelihood (NLL) loss function to perform variational inference, incorporating both the PRB utilization estimates $\mathbf{\hat{x}}_{t+1:t+L}^{(tar)}$ and the corresponding standard deviations $\mathbf{\hat{\sigma}}_{t+1:t+L}^{(tar)}$ predicted by the model (see Algorithm~\ref{alg:contextual-predictor}). For each cell, the model additionally includes information about the antenna type ($\mathbf{m}_i$) and the frequency band where it operates, encoded as one-hot vectors of 15 elements. We implement $ScaleNN$ using a Multi-Layer Perceptron (MLP) of two layers, where the hidden layer consists of 183 units, and we implement the embedding networks $QueryEmbed$ and $KeyEmbed$ as 2-layer MLPs, each producing hidden states ($\mathbf{hq}$, $\mathbf{hk}$) of 6 elements.

We train the Contextual Predictor over 400 epochs, each consisting of 40,000 randomly selected samples, using data from the first 14 days, and an Adam optimizer with a learning rate of $10^{-5}$. Since the training dataset is derived from a production network, it may inherently contain anomalies. These anomalies---which are typically outliers---often result in larger errors and can cause the model to overfit to these cases. To mitigate this effect, we apply gradient clipping ($\|\mathbf{g}\| \leq 0.8$), which limits the magnitude of the gradients during backpropagation. With this, we aim to limit the influence of individual samples in the dataset---including anomalies---and focus the model on capturing the overall trends from the bulk of cells seen during training (\ie learn the normal behavior).

\begin{table}[!t]
\centering
\caption{Comparison of Contextual Predictor models.}
\label{tab:comp-predictors}
\resizebox{\columnwidth}{!}{
\begin{tabular}{@{\extracolsep{4pt}}lccccc}
& \multicolumn{2}{c}{\textbf{Norm. MAE}} & \multicolumn{2}{c}{\textbf{\textit{R\textsuperscript{2}}}} & \multirow{2}{*}{\textbf{\shortstack{Interpretable\\Context}}} \\
\cline{2-3} \cline{4-5} \\
\textbf{Model} & \textbf{Train} & \textbf{Validation} & \textbf{Train} & \textbf{Validation} & \\
\toprule
\textbf{Our model} & \textbf{0.169} & \textbf{0.158} & \textbf{0.930} & \textbf{0.931} & \checkmark \\
\hspace{2mm}- w/o Auto-Scaler & 0.215  & 0.206 & 0.888 & 0.889 & \checkmark \\
\hspace{2mm}- w/o Linear Combination & 0.191 & 0.182  & 0.909 & 0.909 & \xmark \\
\bottomrule
\end{tabular}
}
\end{table}

Table~\ref{tab:comp-predictors} (top row) shows the prediction accuracy of the model in terms of both absolute error and the coefficient of determination ($R^2$). We put the results into context by normalizing the Mean Absolute Error (MAE) by the mean value of PRB utilization seen in the training dataset (\ie \textit{Norm. MAE}). The results on both the training and validation datasets exhibit similar values, suggesting that the model generalizes effectively when applied to cells from a different deployment area not included in the training phase.

We further evaluate two model variants of the proposed Contextual Predictor:\\
\noindent$\bullet$ \textbf{Model without Auto-Scaler:} This variant omits the Auto-Scaler module (see Algorithm~\ref{alg:contextual-predictor}). Instead, it processes the cell-level data $c_i \in \mathcal{N}$ directly as input to embedding functions in the Self-Attention module. Lastly, it computes predictions $\hat{x}_{t+1:t+L}^{(tar)}$ and standard deviations $\hat{\sigma}_{t+1:t+L}^{(tar)}$ in the same manner as the proposed Contextual Predictor. \\
\noindent$\bullet$ \textbf{Model without Linear Combination:} We replace the weighted linear combination in the Readout module of our Contextual Predictor with a Neural Network (NN). Similar to traditional GATs~\cite{velivckovic2017graph}, the model combines the hidden states of the neighbors considering the attention coefficients. Then, the readout NN (implemented as a 2-layer MLP) processes the aggregated hidden states along with additional embeddings of the time series of neighboring cells during the Prediction Window $\hat{x}_{t+1:t+L}^{(neigh)}$. Note that due to the non-linear nature of these embeddings, the attention coefficients and the resulting context produced by the model are not directly interpretable.

The results (Table~\ref{tab:comp-predictors}) show that the Auto-Scaler module improves the model's accuracy across both evaluated metrics. Additionally, the use of a linear combination in the Readout module does not incur a performance penalty. Instead, it makes the context of the model interpretable and contributes to reducing the false alarm rate, as detailed in Sec.~\ref{subsubsec:anomaly-detector}.

\subsection{Anomaly Detector calibration}\label{subsec:calibration}

\begin{figure}[!t]
\begin{minipage}[b]{0.47\columnwidth}
    \centering
    \includegraphics[width=\linewidth]{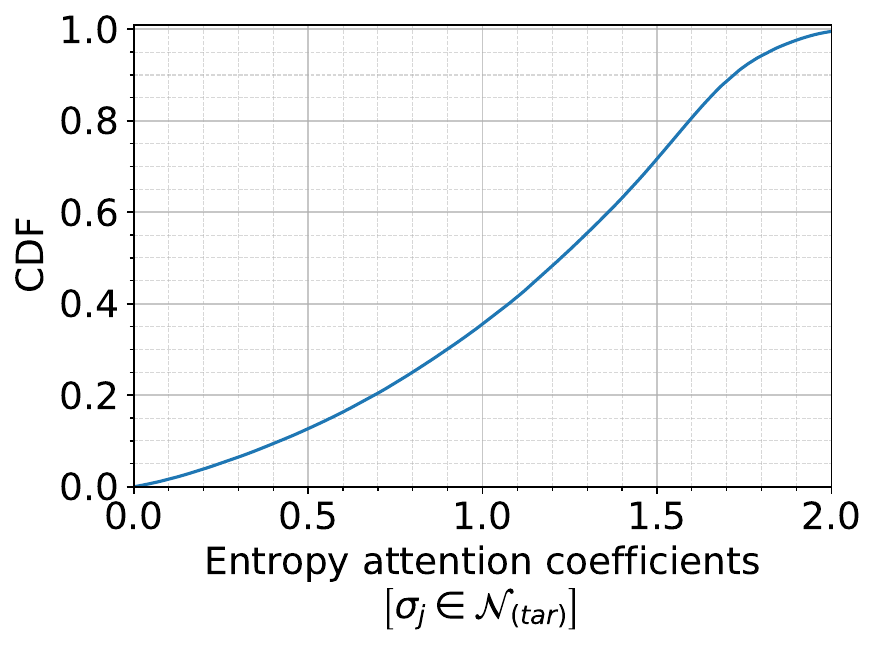}
    \vspace{-0.7cm}
    \caption{Distribution of the entropy of attention coefficients.}
    \label{fig:entropy-att-coef}
\end{minipage}%
\hfil
\begin{minipage}[b]{0.47\columnwidth}
    \centering
    \includegraphics[width=\linewidth]{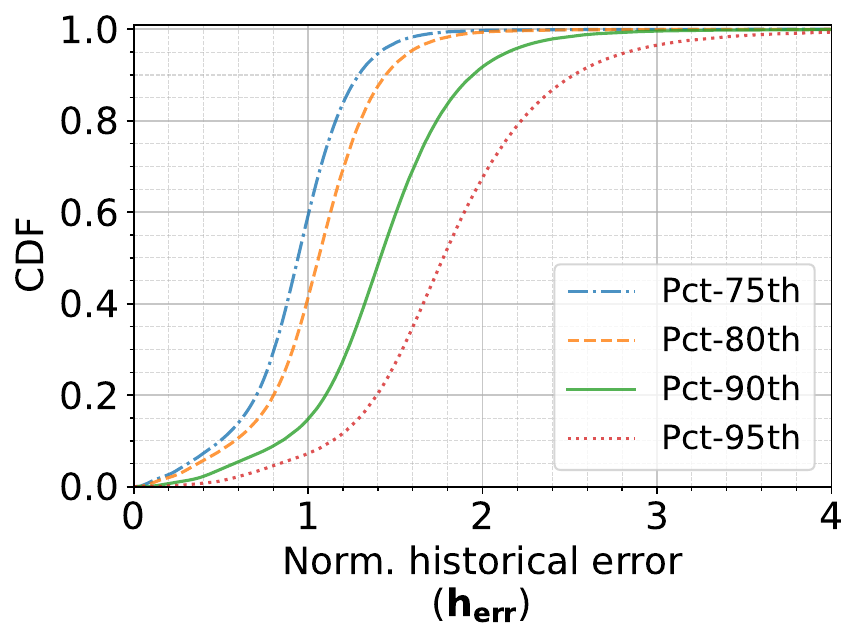}
    \vspace{-0.7cm}
    \caption{Distribution of the normalized historical error.}
    \label{fig:hist-context}
\end{minipage}

\begin{minipage}[b]{\columnwidth}
    \centering
    \vspace{1em} 
    \includegraphics[width=0.47\columnwidth]{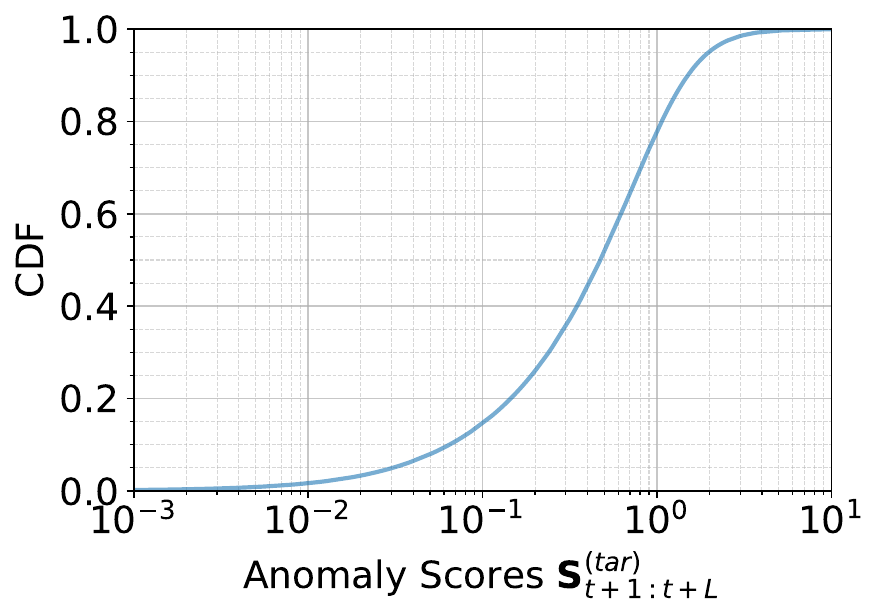}
    \vspace{-0.3cm}
    \caption{Distribution of the anomaly scores.}
    \label{fig:cdf-pred-error}
\end{minipage}
\end{figure}

After training the Contextual Predictor, we configure the parameters of the Anomaly Detector, namely ($i$)~the entropy threshold of the attention coefficients $\lambda$, ($ii$)~the parameters related to the historical error ($P$ and $\gamma$), and ($iii$)~the anomaly detection threshold $\delta$, which are defined in Sec.~\ref{subsubsec:anomaly-detector}.

Figure~\ref{fig:entropy-att-coef} illustrates the entropy of the attention coefficients for all samples in the validation dataset. Likewise, Figure~\ref{fig:hist-context} presents the historical errors per sample, specifically reporting the 75th, 80th, 90th, and 95th percentiles computed over the Context Window ($T$=167 hours). Based on our experimentation and conversations with an operations team of the MNO, we set a threshold of $\lambda$=0.7, and for the historical error, we take as a reference the percentile $P$=90 and set $\gamma$=2. Additionally, Figure~\ref{fig:cdf-pred-error} shows the anomaly scores produced by the Anomaly Detector. Based on that, we establish a sensitivity threshold of $\delta$=5. 

With these parameters, we apply the Anomaly Detector over the test dataset and observe that 72.7\% of samples satisfy the filters of entropy and historical errors, and from these samples we detect 0.243\% of data points over the threshold $\delta$ (\ie~anomalies).

\subsection{Evaluation of the anomalies}\label{subsec:evaluation-anomalies}

\begin{figure}[!t]
    \centering
  \subfloat[Class \#1\label{fig:class-1}]{%
       \includegraphics[width=1.0\linewidth]{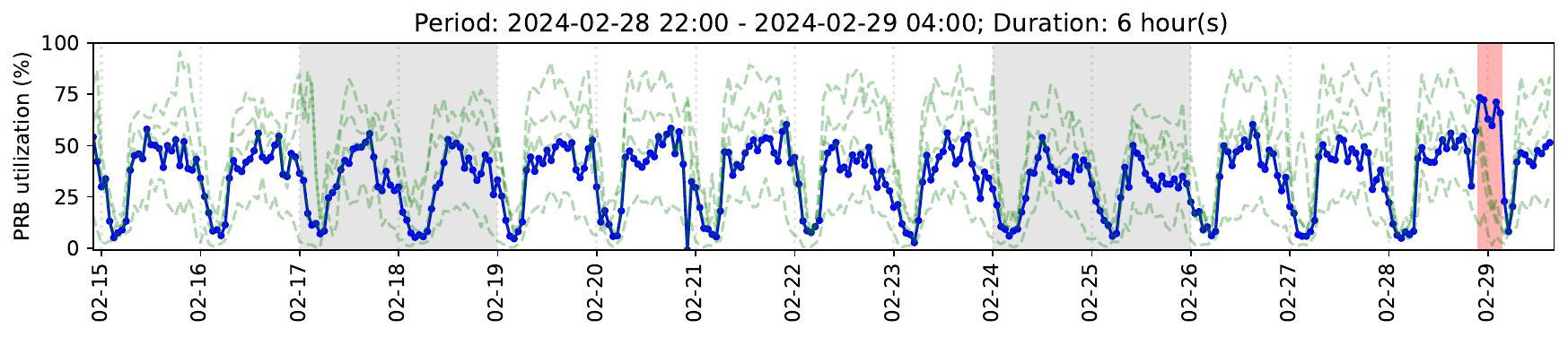}} 
    \hfill
  \subfloat[Class \#2\label{fig:class-2}]{%
        \includegraphics[width=1.0\linewidth]{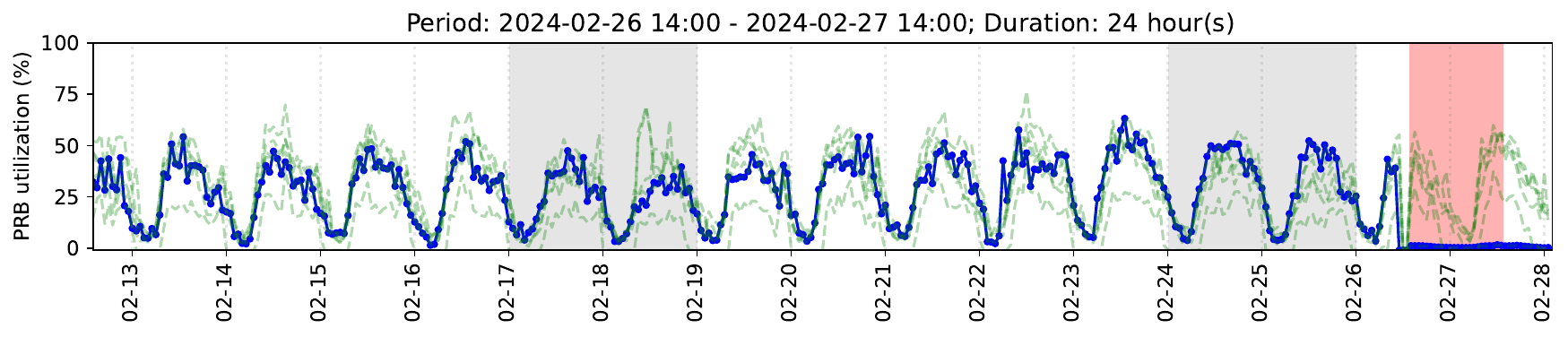}}
  \caption{Examples of anomalies. Solid blue lines show the PRB utilization of the affected cell. Dashed green lines show the utilization of the other cells in the same sector. Red shades mark anomalous periods; gray shades indicate the weekends.}
\end{figure}

In this section, we assess the anomalies detected by \anemon. Note that samples in our dataset are generated over hourly time windows, while the Anomaly Detector produces anomaly scores $\mathbf{S}_{t+1:t+L}^{(tar)}$ for each sample over the entire Prediction Window ($L$=24 hours). Consequently, for each hourly data point our solution generates 24 anomaly scores. To consolidate them into a single label per cell and per hour, we consider a data point to be anomalous if any of its corresponding scores is above the threshold $\delta$. Then, we group consecutive anomalous hours into time periods and treat these periods as single cell-level anomalies spanning specific durations.

We further analyze the types of anomalies detected by performing a manual inspection. Typically, the most critical anomalies are those that persist for extended periods. Based on this, we focus on the subset of anomalies lasting more than six hours, which results in 74 anomalous periods distributed across 63 cells.

To inspect these anomalies, we follow a procedure informed by feedback from an operations team of the MNO: during the anomalous hours, we examine the evolution of the PRB utilization of all cells in the same sector as the target cell. This analysis helps determine whether the variation in utilization in the affected cell can be explained by the behavior of the other cells covering the same geographical area (\eg due to temporary cell outages in the sector). Based on that, we define the following two main classes of anomalies:

\noindent \textbf{\#1) Unexpected variation in cell utilization:} It includes cases where the affected cell shows a temporary deviation in utilization, while the other cells in the same sector remain within expected ranges based on their past temporal evolution (see an example in Figure~\ref{fig:class-1}). This indicates that the variation cannot be explained by the transfer of users to/from the affected cell. Such anomalies often point to issues requiring intervention from the operations team, such as misconfigurations, hardware failures, or interference.

\noindent \textbf{\#2) Extreme behavior shifts:} This class includes cases where the utilization distribution of a cell changes drastically for an extended period, either from relatively high levels to very low levels, or vice versa. It also covers cases where a cell transitions between active and inactive states for a long period (example in Figure~\ref{fig:class-2}). Although such anomalies can be easily detected by alternative methods---such as adaptive thresholds---we include them in our evaluation to validate the capability of \anemon to detect these extreme cases.

Table~\ref{tab:classification-anomalies} shows the number of anomalies detected in each class. Notably, a significant proportion of anomalies correspond to Class \#1 (45.94\%), which are of particular interest to operations teams due to their frequent need for intervention. Following the manual inspection, we observe that \mbox{non-classified} anomalies ($N/A$) mostly exhibit unexpected behaviors; however they require further investigation, including an in-depth analysis of the status of the fronthaul and midhaul connectivity, as well as supplementary monitoring data to determine if they are related to relevant incidents. The investigation of these complex anomalies requires combining multiple data sources from different engineering teams within the MNO. Given this complexity, we look forward to addressing the validation of these anomalies in future work.

\begin{table}[!t]
\centering
\caption{Classification of long-lasting anomalies of 6+ hours.}
\label{tab:classification-anomalies}
\begin{tabular}{@{\extracolsep{4pt}}ccc}
\textbf{Class \#1} & \textbf{Class \#2} & \textbf{N/A} \\
\toprule
\textbf{34} (45.95\%) & \textbf{14} (18.91\%) & \textbf{26} (35.14\%) \\
\bottomrule
\end{tabular}
\end{table}

\subsection{Robustness on a massive event}\label{subsec:robustness-evaluation}

We test the capability of \anemon to adapt to extreme mobility patterns on a popular event, where we observe a significant shift in the cell-level utilization distribution. Note that these types of events should not trigger anomaly alarms from \anemon, as they are not related to network issues but are instead caused by external factors (\ie the aglomeration of connected users in the area of study). Particularly, we take as a reference a popular football match in a well-known football stadium in the studied metropolitan area, which has a capacity of $>$80k spectators. Note that this area was not included in the training dataset of the Contextual Predictor. Since the event took place approximately during 9pm-11pm, we apply our anomaly detection solution over a broader time window (3:00pm-11:59pm) to capture the complete mobility phenomenon. As a result, we observe that\mbox{---despite} the notable utilization peak during the match---our solution does \textit{not} detect any anomaly in the surrounding area of the stadium (204 cells). 
This is because the GNN-based model calibrates its predictions based on the behavior of neighboring cells, which are also affected by the same spatio-temporal mobility pattern. This evidences the robustness of the proposed solution to be unaffected by external mobility factors.

\section{Conclusions}
In this paper we have presented \anemon, a contextual anomaly detection monitor tailored to focus exclusively on network-related incidents within the RAN. Our solution\mbox{---based} on a custom GNN-based model---captures time series of KPIs in cells and automatically calibrates detections based on the behavior of neighboring cells (\ie local context). Unlike existing RAN anomaly detection methods, our approach enables the detection of anomalies that are unaffected by external mobility factors and allows using a single Deep Learning model that we can apply over large deployment areas.

We have evaluated \anemon using data from a major MNO in a European metropolitan area. As a result, we have defined two primary classes of long-lasting anomalies (6+ hours). From these classes, we have identified a group of anomalies (45.95\%) that is particularly relevant from an operational perspective. As future work, we plan to further investigate known past incidents in the MNO's network and classify anomalies at a finer level of granularity based on their root cause.

\section*{ACKNOWLEDGMENT}
This work has been supported by ($i$) the Spanish Ministry of Economic Affairs and Digital Transformation and the European Union – NextGeneration EU, in the framework of the Recovery Plan, Transformation and Resilience (PRTR) through the UNICO I+D 5G SORUS-RIS project, ref. number TSI-063000-2021-138, and ($ii$) the European Commission through Grant No. 101139270 (ORIGAMI).

\balance
\bibliographystyle{ieeetr}
\bibliography{references}

\end{document}